\documentclass[10pt]{bmc_article}

\usepackage{cite} 
\usepackage{url}  
\usepackage{ifthen}  
\usepackage{multicol}   
\usepackage{multirow}
\usepackage[utf8]{inputenc} 
\usepackage{graphicx}
\newboolean{publ}

\newenvironment{bmcformat}{\begin{raggedright}\baselineskip20pt\sloppy\setboolean{publ}{false}}{\end{raggedright}\baselineskip20pt\sloppy}

\begin{document}
\begin{bmcformat}
\title{CompostBin: A DNA composition-based algorithm for binning environmental shotgun reads}

\author{
        Sourav Chatterji$^{1}$%
        \email{Sourav Chatterji - schatterji@ucdavis.edu}%
        \and%
        Ichitaro Yamazaki$^{2}$%
        \email{Ichitaro Yamazaki - yamazaki@cs.ucdavis.edu}%
        \and%
        Zhaojun Bai$^{2}$%
        \email{Zhaojun Bai - bai@cs.ucdavis.edu}%
        \and%
        Jonathan A Eisen\correspondingauthor$^{1,3,4}$%
        \email{Jonathan A Eisen - jaeisen@ucdavis.edu}%
      }

\address{%
  \iid(1)UC Davis Genome Center\\%
  \iid(2)Department of Computer Science, UC Davis\\%
  \iid(3)Section of Evolution and Ecology, UC Davis\\%
  \iid(4)Department of Medical Microbiology and Immunology, UC Davis%
}%

\maketitle

\begin{abstract}

A major hindrance to studies of microbial diversity has been that the vast majority of microbes cannot be cultured in the laboratory and thus are not amenable to traditional methods of characterization. Environmental shotgun sequencing (ESS) overcomes this hurdle by sequencing the DNA from the organisms present in a microbial community. The interpretation of this metagenomic data can be greatly facilitated by associating every sequence read with its source organism. We report the development of CompostBin, a DNA composition-based algorithm for analyzing metagenomic sequence reads and distributing them into taxon-specific bins. Unlike previous methods that seek to bin assembled contigs and often require training on known reference genomes, CompostBin has the ability to accurately bin raw sequence reads without need for assembly or training. It applies principal component analysis to project the data into an informative lower-dimensional space, and then uses the normalized cut clustering algorithm on this filtered data set to classify sequences into taxon-specific bins. We demonstrate the algorithm's accuracy on a variety of simulated data sets and on one metagenomic data set with known species assignments. CompostBin is a work in progress, with several refinements of the algorithm planned for the future.
       
\end{abstract}

\ifthenelse{\boolean{publ}}{\begin{multicols}{2}}{}

\section*{Background}
Microbes are ubiquitous organisms that play pivotal roles in the earth's bio-geochemical cycles. Their most visible effects on human well-being arise through their roles as mutualistic symbionts and hazardous pathogens. The study of microbes is crucial to our understanding of the earth's life processes and human health. Most of our knowledge about microbes has been obtained through the study of organisms cultured in artificial media in the laboratory. Although this approach has provided profound biological insights, it is inadequate for studying the structure and function of many microbial communities. One obstacle has been that the vast majority of microbes have not been cultured and may not be culturable  \cite{rappe_majority}.
Even though culture independent methods such as 16S rRNA surveys
\cite{pace1,pace2} have been developed, they are unable to simultaneously answer two fundamental questions: Who is out there? and What are they doing? The application of genome sequencing methods is revolutionizing this field by enabling us for the first time to address those two questions for unculturable microbial communities \cite{venter_saragasso,tyson_nature,gut2006}. 
These techniques, called environmental genomics or metagenomics, study unculturable communities by analyzing the pooled genomes of all the organisms present in the community. The genomic data obtained can be analyzed to make inferences about both who is out there and what they are doing (e.g., \cite{sharp}).
\par
In one specific metagenomic method, {\em {\bf e}nvironmental {\bf s}hotgun {\bf s}equencing} (ESS), DNA pooled from a microbial community is sampled randomly using whole genome shotgun sequencing. Thus, ESS data is made up of sequence reads from multiple species. This adds an additional layer of complexity compared to single-species genome sequencing, as it requires analysis of the metagenomic data in order to associate each sequence read with its source organism. Therefore, a critical first step in many metagenomic analyses is the distribution of reads into taxon-specific bins. 
\par
The difficulty of accurately binning ESS reads from whole genome data remains a significant hurdle in metagenomics. The taxonomic resolution achievable by the analysis depends on both the binning method and the complexity of the community. For instance, binning into species-specific bins can be achieved in low-complexity microbial communities (e.g., the dual-bacterial symbiosis of sharpshooters 
\cite{sharp}). 
However, the problem becomes more difficult in high-complexity communities with hundreds of species, such as the Sargasso Sea \cite{venter_saragasso} and the human distal gut \cite{gut2006}. Because of these difficulties, many metagenomic studies (e.g., \cite{soil_whale}) have resorted to analyzing at the level of the metagenome, essentially treating a microbial community as a bag of genes. This is not a satisfactory solution. Identifying and characterizing individual genomes can provide deeper insight into the structure of the community \cite{sharp}. 
\par
A variety of approaches have been developed for binning: assembly, phylogenetic analysis \cite{bork07}, 
database search \cite{megan}, alignment with reference genome \cite{GOS} and
DNA composition metrics \cite{TETRA,phylopythia} 
Most current binning methods suffer from two major limitations: they require
closely related reference genomes for training/alignment and
they perform poorly on short sequences.
To overcome the second difficulty, almost all current binning methods are applied to assembled contigs. However, most of the current generation assemblers can be confounded by metagenomic data since they implicitly assume that the shotgun data is from a single individual or clone. Therefore, we believe that assembly is risky when binning and that it is necessary to analyze raw sequence reads to get an unbiased look at the data. 
\par
To overcome the above-mentioned disadvantages of other binning methods, we have developed CompostBin, a binning algorithm based on DNA composition. CompostBin can bin raw sequence reads into taxon-specific bins with high accuracy and does not require training on currently available genomes. Like other composition-based methods, it seeks to distinguish different genomes based on their characteristic DNA compositional patterns, termed "signatures." For example, one of the most commonly used DNA metrics measure the frequency of occurrence of  Kmers (oligonucleotides of length $K$)in a DNA sequence. Kmer frequencies have been used to distinguish between organisms since the 1960s \cite{kornberg}. With the explosion of available genomic data in the 1990s, Karlin and colleagues were able to establish that the relative abundances of various dinucleotide sequences (the dinucelotides odds ratio) is a genomic signature \cite{karlin}.
Subsequently, taxon-specific biases were also found in the frequencies of Kmers with lengths of four or more, leading to the use of a wide variety of methods exploiting this bias as a signature \cite{CGR,TUD,SOM,TETRA,phylopythia,salzberg}.
\par
Unfortunately, many composition-based binning algorithms do not perform well on short fragments. Poor performance in shorter fragments is caused by the noise associated with the high dimensionality of the feature space. When measuring the frequency of Kmers, the feature vector has $4^K$  dimensions (associated with measuring the frequencies of $4^K$ possible oligonucleotides of length $K$). Thus, for instance, if one looks at the frequency of hexamers in 2kb fragments, the dimensionality of the feature space is twice the length of the sequenced fragments. 
\par
CompostBin employs a new approach to deal with the noise arising from the high dimensionality of the feature vector. Instead of treating all components of the noisy feature space equally, we extract the most "important" directions and use these  components for distinguishing between taxa. The technique employed is Principal Component Analysis (PCA) \cite{PCA}, a multivariate analysis method previously applied in diverse biological areas ranging from ecology \cite{PCA-ecology} to  codon usage in genes \cite{PCA-codon} and  even visualization of metagenomic binning results \cite{tanja}. The normalized cut clustering algorithm used to cluster sequences into taxon-specific bins is further guided by information from phylogenetic markers. We tested CompostBin on a wide variety of data sets and demonstrated that it is highly accurate in separating sequences into taxon-specific bins, even when processing raw reads of short sequences.

\section*{Results}
CompostBin was coded in C and Matlab on a 64bit Linux Machine. It is publicly available for download from the Eisen Lab website. As shown in the overview in Figure 1, CompostBin extracts the "principal components" of the DNA composition data and then uses PCA to project that data into a lower-dimensional space for further analysis. As shown in Figure 2, the algorithm can distinguish sequences from various species using just these first three principal components. Next, CompostBin uses the normalized cut clustering algorithm \cite{ncut} to segment the data set into taxon-specific bins. Since the accuracy of phylogenetic assignment for reads containing phylogenetic marker genes is very high \cite{bork07}, we devised a semi-supervised approach which uses the phylogenetic information to guide the clustering algorithm. Simulated data sets were designed to evaluate the accuracy of CompostBin in binning metagenomic data sets of low and medium complexity. Additionally, we tested the data set on environmental shotgun reads from the gut of a glassy-winged sharpshooter \cite{sharp}. Details of the test data sets and CompostBin's performance are provided in the next two sections. 
\subsection*{Test Data Sets}
Metagenomics being a relatively new field, standard data sets for testing binning algorithms have not yet been developed. One obstacle to their development has been that the "true" solution is still unknown for the sequence data generated by most metagenomic studies. To test the accuracy of a binning algorithm, one can instead simulate the shotgun sequences that would be obtained from a combination of organisms of known genome sequences. We used ReadSim \cite{readsim}, a publicly available program, to simulate Sanger sequences from known genomes. The sequence reads from multiple genomes were pooled to simulate the challenges of metagenomic sequencing. When designing our simulated data sets, we took into account several variables that affect the difficulty of binning: the number of species in the sample, their relative abundance, their phylogenetic diversity, and the differences in GC content between genomes. 
\par
Since environmental shotgun data is influenced by factors that may not be reflected in simulation experiments, we also tested CompostBin on a publicly available metagenomic data set whose solution is well accepted. Data Set R1 contains sequence reads obtained from gut bacteriocytes of the glassy-winged sharpshooter, {\em Homalodisca coagulata}. The data sets used for testing CompostBin are described in Table 1, and experimental details are provided in Methods. 

\subsection*{Performance}
CompostBin's accuracy in classifying reads from the test data sets is reported in Table 1. The percentage of misclassified reads is less than $6\%$ in 11 of the 13 data sets. The highest error rates measured were $8.01\%$ for Data Set S3 (sequences from {\em E. coli} and {\em Y. pestis}) and $7.24\%$ for Data Set S5 (sequences from B. anthracis and L. monoytogenes). In both cases, the phylogenetic distance between genomes is comparatively small. However, the results from Data Set S1, which contains sequences from {\em Bacillus halodurans} and {\em Bacillus subtilis}, show that, in some instances, the algorithm can distinguish at the species level with high accuracy. The low error rate for the sharpshooter data set (R1) demonstrates the ability of the algorithm to handle the peculiarities of environmental shotgun data. 
\section*{Discussion}
In this paper, we report the development of a new approach to the taxonomic binning problem associated with the analysis of metagenomic data. Accurate binning is a crucial step in the application of environmental shotgun sequencing to the study of microbial communities. The problem of binning is intertwined with the fundamental questions of genomic signatures. Does the genome of every organism have a unique signature that distinguishes it from the genomes of all others? If so, what is the minimum length DNA sequence required to distinguish between two organisms? Even though it has been demonstrated that DNA-composition metrics such as the dinucleotide odds ratios are genome signatures \cite{karlin}, previous studies have typically worked with long sequences. In this study, we demonstrate that shrewdly analyzed Kmer frequency data from short sequences can also provide a signature. The principal novel aspect of our method is the observation that the high-dimensional Kmer frequency data for short sequences is noisy, and that one can deal with the noise by projecting the data into a carefully chosen lower-dimensional space. This lower-dimensional space is determined by the principal components of the data. In a sense, it, too, is a "genome signature" that can be used to classify even short sequences into taxon-specific bins.
\par
We used the frequencies of hexamers (oligonucleotides of length 6) as the metric for our analysis of short sequences. The choice of hexamers was motivated by both computational and biological rationale. Since the length of the feature vector for analyzing Kmers is $O(4^K)$, both the memory and the CPU requirements of the algorithm become infeasible for large data sets when $K$ is greater than six. Using hexamers is biologically advantageous in that, being the length of two codons, their frequencies can capture biases in codon usage. Similarly, hexamer frequencies can detect genomic biases resulting from the observed avoidance of specific palindromic words of lengths $4$ and $6$ from genomes due to the presence of restriction enzymes \cite{restriction}. It should be noted that the frequencies of lower-length words are linear combinations of hexamer frequencies. For example: $f (AAAAA) = f (AAAAAA) + f (AAAAAC) + f (AAAAAG) + f (AAAAAT )$. Thus, our PCA-based method implicitly takes into account any biases in the frequencies of lower length words. 
\par
Our method of analysis is based primarily on DNA composition metrics and, like all such methods, it cannot distinguish between organisms unless their DNA compositions are sufficiently divergent. Thus, our method would probably be unable to distinguish between strains of the same species. We believe that an ideal binning algorithm would also utilize additional types of information, such as assembly (depth of coverage and overlap information) and population genetics parameters. We have taken an initial step in this direction by using taxonomic information from phylogenetic markers to guide the clustering algorithm. We intend to develop other hybrid methods in the future. 
\par
An ideal binning system would, like CompostBin, not require training of the algorithm with data from sequenced genomes. This is critical for success when binning environmental shotgun data because more than $99.9\%$ of microbes are currently unculturable and unlikely to be represented in the training data set. Even closely related organisms living in different environments may have divergent genome signatures. For example, Bacillus anthracis and Bacillus subtilis have widely differing GC content and genome signatures. One should also keep in mind that the currently available genomes are not a phylogenetically random sample, but rather are a highly biased collection of biomedically interesting genomes combined with an overabundance of strains of model organisms such as Escherichia coli. 
\par 
CompostBin is a work in progress, with several refinements of the algorithm planned for the future. 
\begin{itemize}
	\item In the analyses reported here, we used PCA as the projection method for choosing the lower-dimensional space. Since PCA misses nonlinear structures of the underlying variables, we plan to look at alternative projection methods such as Projection Pursuit \cite{pp}, ICA \cite{ica}, and kernel PCA \cite{kpca}. 
	\item CompostBin analyzes only the first three principal components in the data set. We plan to explore alternative approaches for choosing the optimal number of principal components (e.g., \cite{pca2}). 
	\item The clustering algorithm employed captures the global geometry of a data set using its $k$-nearest neighbor graph. The highly accurate binning of the data sets reported in this paper was obtained using a fixed value of $k = 6$. However, the optimal value of $k$ may depend on the characteristics of each individual data set. We plan to explore a technique which can automatically determine the optimal $k$ through capture of the global geometry of the input data set. 
	\item The similarity between two connected sequences in the nearest-neighbor graph was measured by the exponential inverse of their normalized Euclidean distance. We plan to explore alternative criteria for sequence similarity which have the potential to improve binning. 
	\item The running time of our program can be improved by developing more efficient data-structures and by utilizing other numerical tools \cite{arpack,TRLAN,PRIME} to compute the principal components of the original data set and the eigenvectors of the similarity matrix. 
	\item We observed that the separate clusters of rRNA genes can be outliers in many archaeal genomes and cause errors in the binning algorithm. Therefore, binning accuracy can be improved in future investigations by removing those genes prior to performing the DNA composition-based analyses. 
	\item We plan to explore other potential applications of our algorithm to the study of genome structure and its variations within a single genome. 
\end{itemize}


\section*{Methods}
\subsection*{Generation of Test Sets}
Genomic sequences of bacterial and archaeal isolate genomes were downloaded from the NCBI GenBank database \cite{genbank}. ReadSim was used to simulate paired-end Sanger sequencing from isolate genomes with an average read length of $1,000$ bp. The reads from various isolates were then combined in ratios corresponding to their relative species abundance in the data set to yield a simulated metagenomic data set of known composition. 
\par
In our experiments, we simulated the sequencing of low- to medium-complexity communities in which the number of species ranged from two to six and their relative abundance ranged from 1:1 to 1:14. We included species of varying degrees of phylogenetic relatedness in order to test the ability of the program to discriminate between sequences at the species, genus, and family levels. The 12 simulated data sets created are described in Table 1.
\par
In addition, we tested the algorithm on a metagenomic data set containing reads obtained from gut bacteriocytes of the glassy-winged sharpshooter. The original study [7] had used phylogenetic markers to classify the sequence reads into three bins: reads from {\em Baumannia cicadellinicola} in Bin 1, reads from {\em Sulcia muelleri} in Bin 2, and reads from the host and miscellaneous unclassified reads in Bin 3. Due to the heterogeneity of Bin 3, the accuracy of the algorithm was tested only on its ability to distinguish between reads from Bin 1 and Bin 2.
\subsection*{The CompostBin Algorithm}
The input to CompostBin consists of raw sequence reads,
along with mate pair information and the taxonomic assignment of reads containing
phylogenetic markers. Either the number of abundant species or
the number of taxonomic groups in the data set is provided to help
the algorithm determine the number of bins in the output. This information can be obtained by 
analyzing the reads containing genes for ribosomal RNA or other marker genes 
\cite{GOS}. In the simulation experiments, the number of bins is set 
to the number of species in the simulation.
\subsubsection*{Feature Extraction by PCA}
Mate pairs are joined together
and treated as a single sequence because they are highly likely to have 
originated
from the same organism. Each sequence being analyzed is initially 
represented as a $4,096$-dimensional feature vector, with each component
denoting the frequency of one of the $4,096$ hexamers. 
As a result, all the sequences are initially represented as
an $N\times 4,096$ feature matrix $A$, where $N$ is the number of sequences being 
analyzed. PCA is then used to decrease the noise inherent in this high-dimensional data set by identifying the principal components of the feature matrix A. 
\par
The PCA algorithm \cite{PCA} filters the noise and removes redundant variables to arrive at a new basis for expressing the data set. Furthermore, by using PCA, we may be able to find new underlying variables which reveal additional details about the mathematical structure of the system. Determining the number of principal components required for analysis is crucial to the success of the algorithm. Too few components and some important information may be lost. Too many components increases the noise in the data unnecessarily. When using PCA to bin sequences, use of just the first three principal components is adequate to separate sequences from different species. Figure 2 shows that for Data Set S5 which contains two alphaproteobacteria with similar GC content, almost complete separation is achieved by using only the first two principal components.
\subsubsection*{Bisection by Normalized Cuts}
The projection of the data matrix $A$ into the first three principal components produces an $N \times 3$ data matrix $A_p$. A clustering algorithm is then applied to $A_p$ to separate the N points into taxon-specific bins. A bisection algorithm is used to bisect a data set into two bins as detailed below. If the data set is to be divided into more than two bins, this algorithm is used recursively. Figure 3 shows pseudocode for the bisection algorithm. Given the projected matrix and phylogenetic markers as inputs, the procedure first computes the weighted graph over the sequences where the edge weights measure the similarity between corresponding sequences. Then, the normalized cut clustering algorithm [22] is employed to bisect the graph such that sequences from the same taxonomic group stay together. 
{\em Computation of Similarity Measure:}
As described earlier, the $4,096$-dimensional feature vector
is projected into the first three principal components, and 
each sequence is represented as a point in $3$-dimensional space. 
The clustering algorithm initially creates a $6$-nearest neighbor 
graph $G(V,E,W)$ 
to capture the structure of the data set. The vertices in $V$ correspond 
to the sequences, and an edge $(v_1,v_2) \in E$ between two 
sequences $v_1$ and $v_2$ exists only if one of the sequences is 
a $6$-nearest neighbor of the other in Euclidean space. 
The nearest-neighbor graph reveals the global relation of the 
data set through this easily-computable local metric~\cite{tene00}.
Each edge between two neighboring sequences $v_1$ and $v_2$ is weighted by 
their similarity $w(v_1,v_2)$, which is defined as the exponential inverse 
of their normalized Euclidean distance:
\[
   w(v_1,v_2) = \left\{\begin{array}{ll}
   e^{-\frac{d(v_1,v_2)}{\alpha}} & \mbox{if } (v_1,v_2) \in E,\\
   0                              & \mbox{otherwise},
   \end{array}\right.
\]
where $d(v_1,v_2)$ is the Euclidean distance between $v_1$ and $v_2$, and 
\[
\alpha=\max_{(v,u) \in E} d(v,u).
\]
\par
{\em Semi-supervision Using Phylogenetic Markers:}
Marker genes, such as the genes that code for ribosomal proteins, are one of the most reliable tools for phylogenetically assigning reads to bins. Since these marker genes appear in only a small fraction of the reads, we used taxonomic information from 31 phylogenetic markers \cite{martin} to improve the clustering algorithm. This taxonomic information is provided to the binning algorithm as a label for each sequence, with each label corresponding to a single taxonomic group. Sequences without a taxonomic assignment are assigned the label "unknown." A semi-supervised approach  can then be employed \cite{kamvar03,yu04} to incorporate this  information into the clustering algorithm.
\par 
Our binning algorithm uses the simplest approach to update the nearest neighbor graph. Two vertices $v_1$ and $v_2$ are connected with the maximum edge weight (i.e., $w(v_1,v_2) = 1$) if the corresponding sequences are from the same taxonomic group, and the edge between $v_1$ and $v_2$ is removed (i.e., $w(v_1,v_2) = 0$) if they are from different groups. 
\par
{\em Normalized Cut and its approximation:}
Given a weighted graph $G(V,E,W)$,
the association between two subsets $X$ and $Y$ of $V$ $W(X,Y)$
is defined as the total weight of the edges connecting $X$ and $Y$:
\[
  W(X,Y) \ = \sum_{x \in X, y \in Y} w(x,y).
\]
The normalized cut algorithm bisects $V$ into two disjoint subsets $U$ and $\bar{U}$ 
such that the association within each cluster is large
while the association between clusters is small, i.e.,
the normalized cut value $NCut$ is minimized, where
\[\label{eq:ncut}
  \mbox{NCut} = \frac{W(U,\bar{U})}{W(U,V)} + \frac{W(U,\bar{U})}{W(\bar{U},V)}.
\]
The minimization of $NCut$ avoids the bias toward small segments,
which results if the cut value is minimized without 
normalization~\cite{wu93}.
Since finding the exact solution to minimize $NCut$
is an NP-hard problem, an approximate solution is computed
using a spectral analysis of the Laplacian matrix of the graph~\cite{ncut}.
To generalize the algorithm for more than two bins, the binning algorithm 
uses PCA and the normalized cut algorithm iteratively, as described below.
\subsubsection*{Generalization to Multiple Bins}
If the data set needs to be divided into more than two bins,
an iterative algorithm is used and sequences in one of the bins
are projected into their first principal components and bisected 
recursively until the required number of bins is obtained.
Figure \ref{fig:binning} shows the pseudocode describing the algorithm. 
A set of bins, $B$ is kept, where each element of $B$ is a set of data points
belonging to the same bin.
The set $B$ is initialized to be the singleton set 
$\{A\}$, where $A$ contains all points in the data set. 
At each subsequent step of the algorithm, 
the bin with the lowest normalized cut value is bisected. The bisection 
continues until either $B$ has the required number of bins or we no longer 
have a good bisection as measured by the normalized cut value.
If none of the bins in $B$ have a small normalized cut value,
the algorithm terminates.
\par
Both the principal components and the normalized cut of $A$ can be computed 
using the Lanczos method~\cite{golub89}
in $\mathcal{O}(N)$ space and $\mathcal{O}(Nm)$ time, where $N$ is the number of sequences in $A$
and $m$ is a small constant representing the number of Lanczos iterations.
By using $kd$-tree~\cite{mount}, a $6$-nearest graph is computed in $\mathcal{O}(N)$ space and $\mathcal{O}(N\log(N))$ time.
Computing and updating the similarity measures takes $\mathcal{O}(N)$ and $\mathcal{O}(l_{\mbox{max}}^2)$ time, respectively,
where $l_{\mbox{max}}$ denotes the maximum number of phylogenetic markers for a particular species in $A$.
\par
In order to separate $A$ into $K$ bins, the bisection algorithm
needs to be called at most $2K-1$ times. Therefore, the running time of
the whole algorithm is bound by $\mathcal{O}(NK(\log(N)))$.
%


\section*{Author contributions}
S.C. and J.E. conceived the high level algorithm and designed the experiments
to test the algorithm's performance. I.Y. and Z.B. were involved in the design 
and analysis of the clustering algorithm. 


\section*{Acknowledgments}

  \ifthenelse{\boolean{publ}}{\small}{}

We thank Lior Pachter, Jonathan Dushoff, Joshua Weitz, 
Dongying Wu, Martin Wu, Amber Hartman, and Jenna Morgan
for their helpful suggestions and comments. S.C. and J.E. were partially 
supported by the Defense Advanced Research Projects Agency under grants 
HR0011-05-1-0057 and FA9550-06-1-0478.
I.Y. and Z.B. were supported in part by the NSF under grant 0313390.









{\ifthenelse{\boolean{publ}}{\footnotesize}{\small}

 \bibliographystyle{bmc_article}  

  \bibliography{pca} }     


\ifthenelse{\boolean{publ}}{\end{multicols}}{}

\newpage





\section*{Figures}

\subsection*{Figure 1 - Overview of the Binning Algorithm}

\begin{figure}[h]
\centering
\includegraphics[width=100mm]{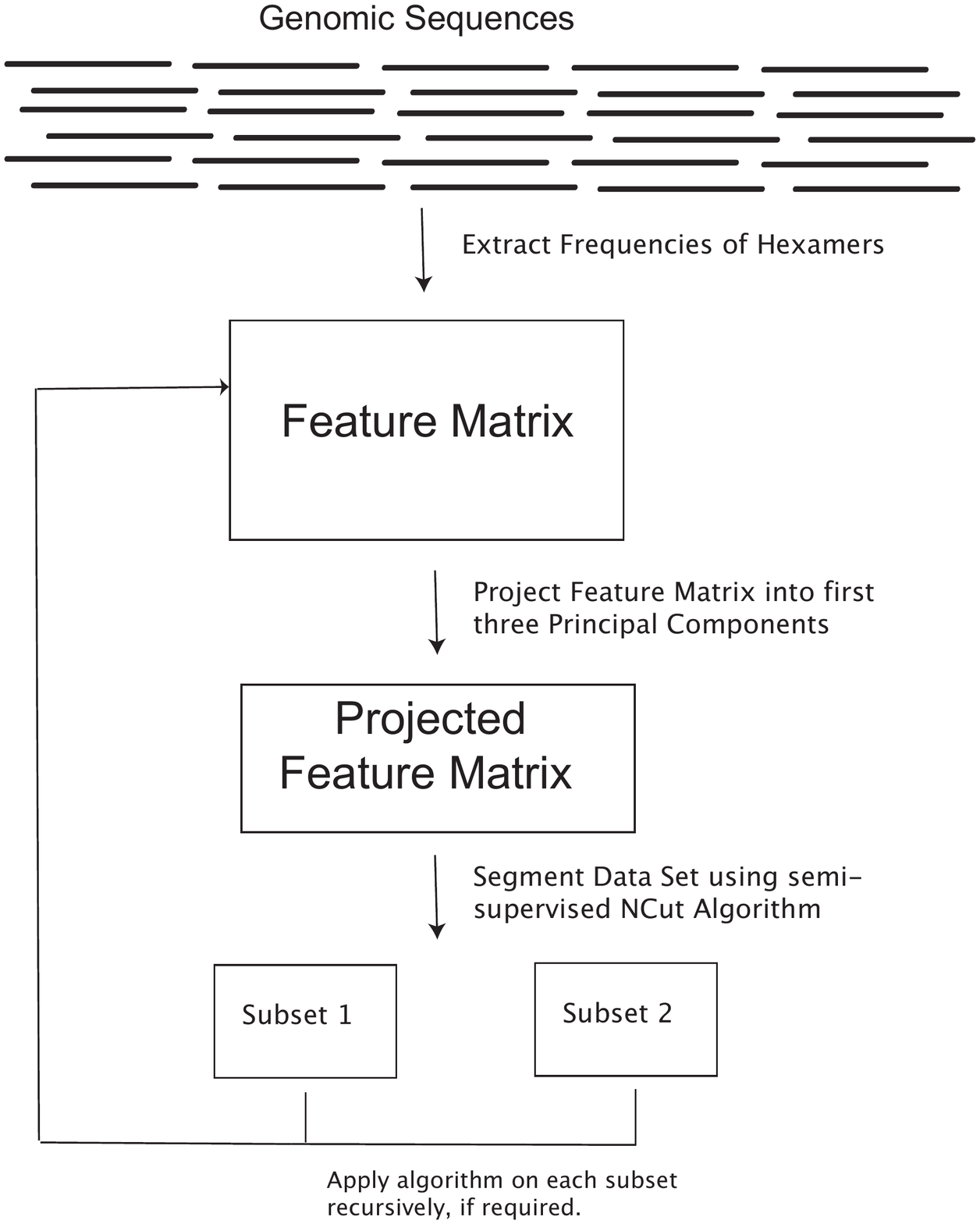}
\caption{High-level overview of the CompostBin algorithm. Each sequence
is represented by a $4,096$-length feature vector, where each component of the
vector represents the frequency of one of $4,096$ hexamers. Thus, $N$ 
sequences are initially represented as a $4,096 \times N$ feature matrix.
Principal Component Analysis is used to project the data into a lower-dimensional 
space. A semi-supervised normalized cut algorithm is used to
segment the data set into two subsets. The algorithm is applied iteratively 
on the subsets to obtain the desired number of bins.
}
\label{fig:overview}
\end{figure}
\newpage

\subsection*{Figure 2 - Separation of sequences by PCA}

\begin{figure}[h]
\centering
\includegraphics[width=180mm]{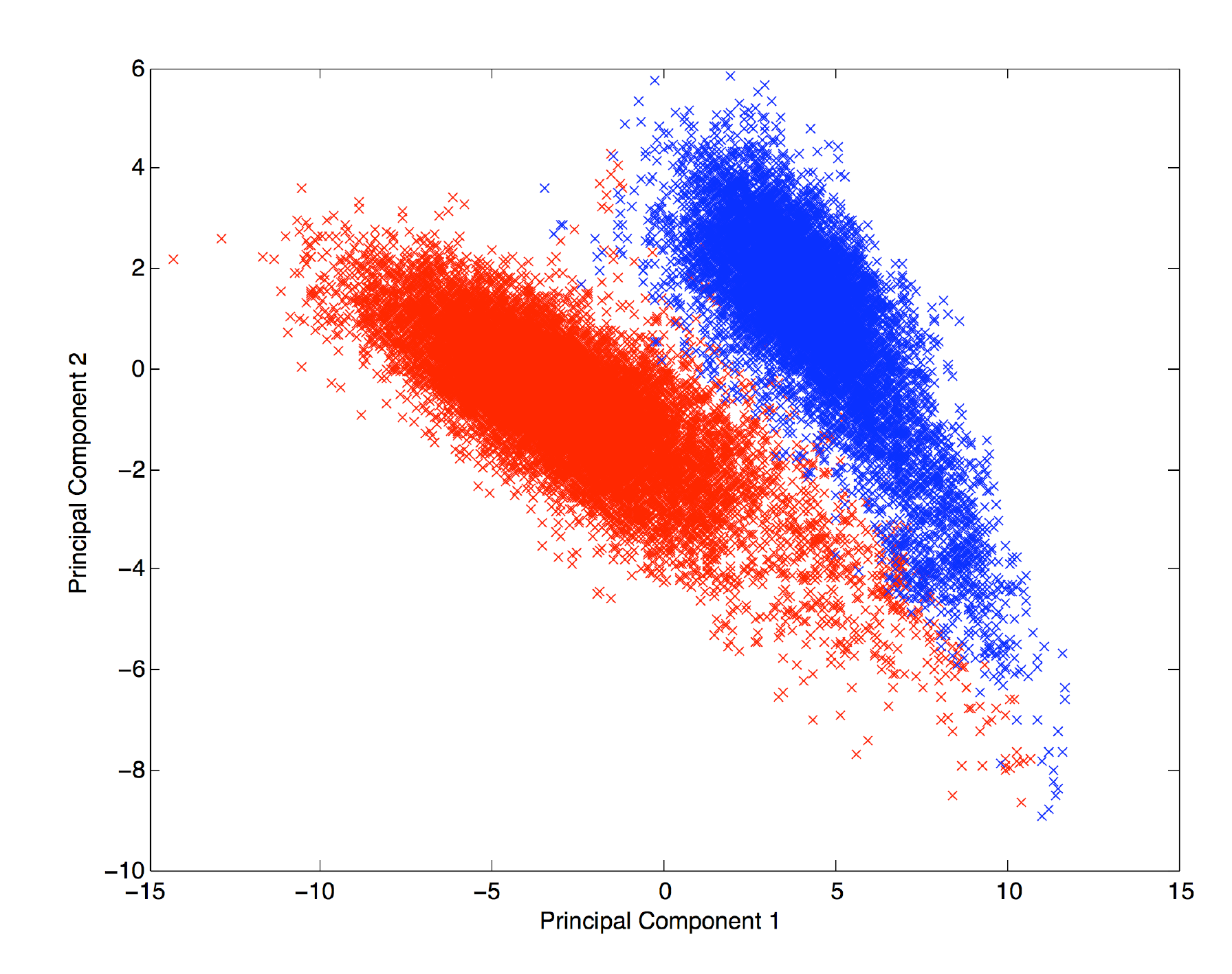}
\caption{Figure illustrating the separation of sequences according to 
species by using just the first few principal components of the data.
This data set contains sequences from two
alphaproteobacteria, {\em Gluconobacter oxydans}
and {\em Rhodospirillum rubrum}, which have GC content of
$0.65$ and $0.61$, respectively. The data set is projected into the 
first two principal components. Sequences from {\em Gluconobacter oxydans}
are represented in red, whereas sequences from {\em Rhodospirillum rubrum} 
are represented in blue. }
\label{fig:separation}
\end{figure}

\newpage
 \subsection*{Figure 3 - The Bisection Algorithm}
\begin{figure}[htp]
\centering
\includegraphics{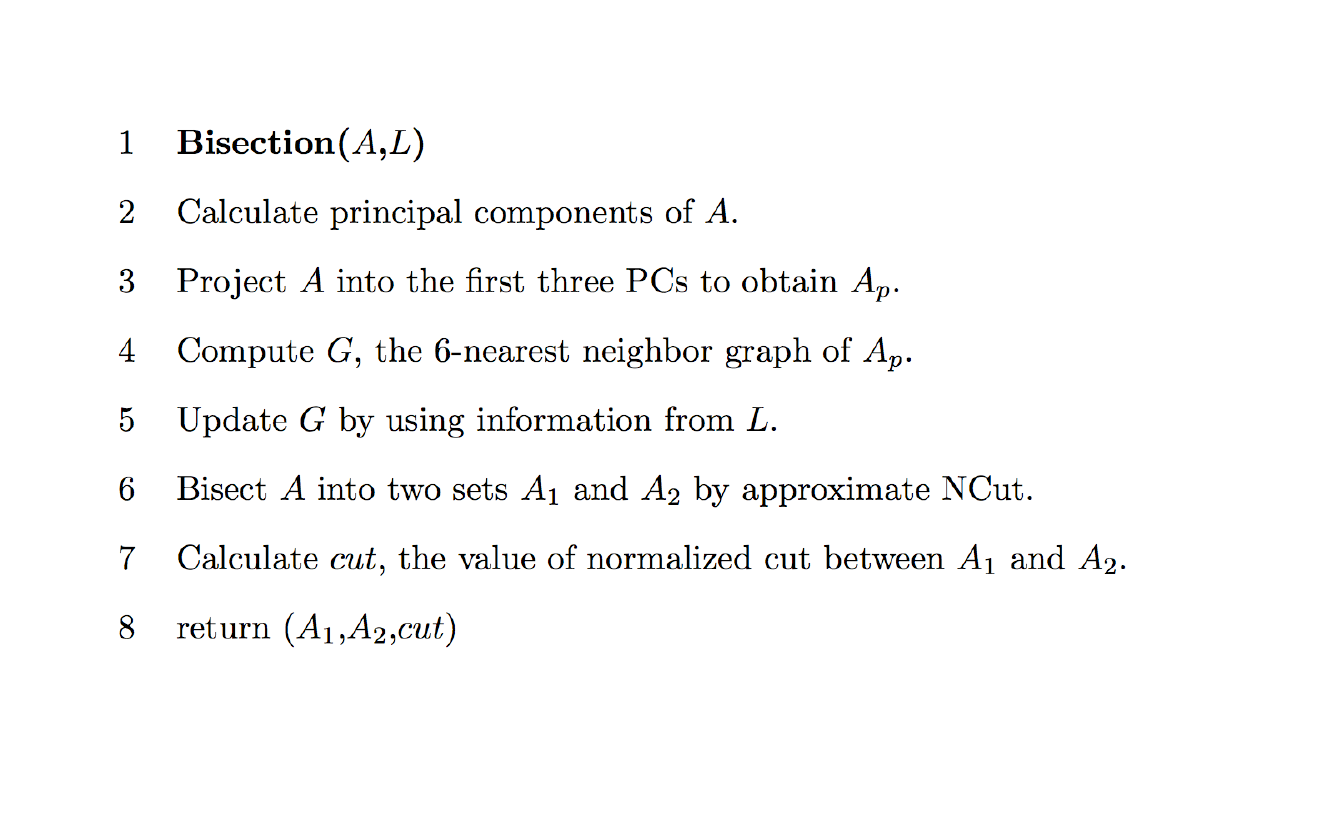}
\caption{
Pseudocode describing the bisection algorithm used to bisect a data set
into two taxon-specific subsets. $A$ is the feature matrix and 
$L$ contains the labeling information for $A$.  
This procedure is used iteratively by the binning algorithm described
in \ref{fig:binning}.}
\label{fig:bisection}
\end{figure}

\newpage

\subsection*{Figure 4 - The Binning Algorithm}

\begin{figure}[htp]
\centering
\includegraphics{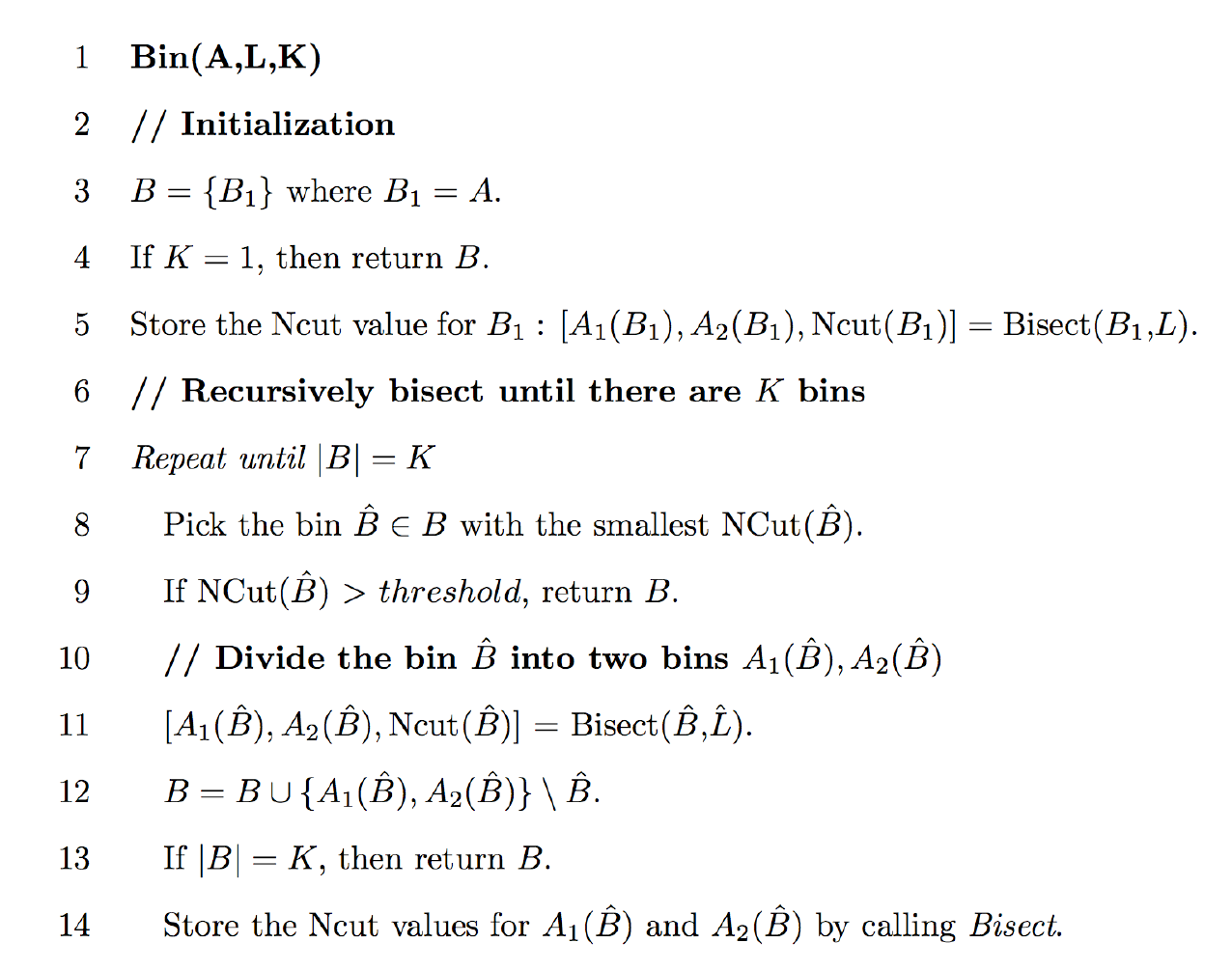}
\caption{Pseudocode describing the iterative PCA and the normalized cut
algorithm used for binning. $A$ is the $N \times 4,096$ feature matrix, with
each $4,096$-length feature vector representing a sequence. 
$L$ contains labeling information obtained from phylogenetic markers,
and $K$ is the the desired number of bins. Lines in bold starting with ``//''
contain comments intended to help understand the code. Note that the 
calls to {\tt Bisect} in Line $11$ can be avoided at the cost of extra 
memory if one stores the optimal cut for each set in $B$ during the calls to
{\tt Bisect} in Lines $5$ and $14$.}
\label{fig:binning}
\end{figure}


\newpage






\section*{Tables}

  \subsection*{Table 1 - Test Data Sets and Binning Accuracy}
Table describing the simulated and real data sets used to test the 
binning algorithm.
Each data set is assigned a unique ID for reference. IDs of
simulated data sets start with {\em S} and  IDs of experimental data sets
start with {\em R}. The GC content of each 
species' genome is listed in squared-brackets and can be used
for assessing the diversity of DNA composition. The taxonomic
levels are obtained from IMG\cite{IMG} and can be used for assessing the 
phylogenetic diversity. The error rate of the binning algorithm
on each test set is shown in the last column. 
\\


      \begin{tabular}{|l|l|c|c|c|}\hline 
	 \multirow{2}{*}{\bf ID}& \multirow{2}{*}{\bf Species} & \multirow{2}{*}{\bf Ratio} & {\bf Taxonomic} & \multirow{2}{*}{\bf Error}\\ 
	 & & &{\bf Differences} & \\\hline
	S1 & {\em Bacillus halodurans} [0.44] \& {\em Bacillus subtilis} [0.44] & 1:1& Species & 5.74\% \\ \hline
    
	S2 & {\em Gluconobacter oxydans} [0.61] \& {\em Granulobacter bethesdensis} [0.59] & 1:1 & Genus & 3.69\% \\ \hline

	S3 & {\em Escherichia coli} [0.51] \& {\em Yersinia pestis} [0.48] & 1:1  &  Genus & 8.01\% \\ \hline  

	S4 & {\em Rhodopirellula baltica} [0.55] \& {\em Blastopirellula marina} [0.57] & 1:1 & Genus & 1.98\% \\ \hline  
	  
	S5  & {\em Bacillus anthracis} [0.35] \& {\em Listeria monocytogenes} [0.38] & 1:2 & Family & 7.24\% \\\hline
	
	\multirow{2}{*}{S6}  & {\em Methanocaldococcus jannaschii} [0.31] \& & \multirow{2}{*}{1:1} & \multirow{2}{*}{Family} & \multirow{2}{*}{0.56\%} \\
    & {\em Methanococcus mariplaudis} [0.33] & & &\\\hline

	S7 & {\em  Thermofilum pendens} [0.58] \& {\em Pyrobaculum aerophilum[0.51]} & 1:1 &  Family & 0.21\%\\ \hline

	S8 & {\em Gluconobacter oxydans} [0.61] \& {\em Rhodospirillum rubrum} [0.65] & 1:1 & Order & 1.15\% \\ \hline

	\multirow{2}{*}{S9} & {\em Gluconobacter oxydans} [0.61], {\em Granulobacter bethesdensis} [0.59], \& & \multirow{2}{*}{1:1:8} & Family & \multirow{2}{*}{2.28\%}\\
	& {\em Nitrobacter hamburgensis} [0.62] & & Order &\\ \hline

	\multirow{2}{*}{S10} & {\em Escherichia coli} [0.51], {\em Pseudomonas putida} [0.62], \&  & \multirow{2}{*}{1:1:8}  &  Order &\multirow{2}{*}{1.73\%}\\
	& {\em Bacillus anthracis} [0.35]& &Phylum &\\\hline

	\multirow{2}{*}{S11} & {\em Gluconobacter oxydans} [0.61], {\em Granulobacter bethesdensis} [0.59], & \multirow{2}{*}{1:1:4:4} & Family & \multirow{2}{*}{5.28\%}\\
	& {\em Nitrobacter hamburgensis} [0.62], \& {\em Rhodospirillum rubrum} [0.65]& & Order &\\ \hline

	 \multirow{3}{*}{S12} & {\em Escherichia coli} [0.51], {\em Pseudomonas putida} [0.62], & 1:1:  &  Species, Order &\multirow{3}{*}{3.35\%}\\
	& {\em Thermofilum pendens} [0.58], {\em Pyrobaculum aerophilum} [0.51], & 1:1:& Family, Phylum &\\
	& {\em Bacillus anthracis} [0.35], \& {\em Bacillus subtilis} [0.44] & 2:14&Kingdom &\\\hline
	 
	 R1 & Glassy-winged sharpshooter endosymbionts & - & - & 5.9\% \\ \hline
	 

      \end{tabular}\label{table:sim}


\newpage



%
\section*{Additional Files}
  \subsection*{Additional file 1 --- CompostBin Code}
%
%
%
File 1, in tar gunzipped format contains the CompostBin source code in C/Matlab.

\subsection*{Additional file 2 --- Test Data Sets}
%
File 2, in tar gunzipped format contains the data sets that was used to test CompostBin's performance.

\end{bmcformat}

\end{document}